# TOWARD ARTIFICIAL EMPATHY FOR HUMAN-CENTERED DESIGN: A FRAMEWORK

**Qihao Zhu**
Data-Driven Innovation Lab
Singapore University of Technology and Design
qihao_zhu@mymail.sutd.edu.sg

**Jianxi Luo**
Data-Driven Innovation Lab
Singapore University of Technology and Design
luo@sutd.edu.sg

May 13, 2023

## ABSTRACT

In the early stages of the design process, designers explore opportunities by discovering unmet needs and developing innovative concepts as potential solutions. From a human-centered design perspective, designers must develop empathy with people to truly understand their needs. However, developing empathy is a complex and subjective process that relies heavily on the designer's empathic capability. Therefore, the development of empathic understanding is intuitive, and the discovery of underlying needs is often serendipitous. This paper aims to provide insights from artificial intelligence research to indicate the future direction of AI-driven human-centered design, taking into account the essential role of empathy. Specifically, we conduct an interdisciplinary investigation of research areas such as data-driven user studies, empathic understanding development, and artificial empathy. Based on this foundation, we discuss the role that artificial empathy can play in human-centered design and propose an artificial empathy framework for human-centered design. Building on the mechanisms behind empathy and insights from empathic design research, the framework aims to break down the rather complex and subjective concept of empathy into components and modules that can potentially be modeled computationally. Furthermore, we discuss the expected benefits of developing such systems and identify current research gaps to encourage future research efforts.

## 1 Background and Introduction

Early-stage conceptual design is crucial for innovation and focuses on discovering unmet human needs and generating design concepts to address such needs. Human-social approaches such as design thinking [1] and ideation techniques such as TRIZ [2] and design heuristics [3] are often used to support and guide such design activities. During the era of data-driven innovation [4], machine learning and data-driven approaches have been growingly adopted to discover and evaluate design opportunities and generate and evaluate design concepts by drawing information, knowledge, and inspiration from data [5-7]. Recent contributions also showed the capability of cutting-edge data-driven artificial intelligence (AI) such as generative pretrained transformers (GPT) for automatic design concept generation [8, 9].

Generating new design concepts with extended digital knowledge sources and improved creative thinking capability with the aid of AI will contribute to increased quantity and novelty of the generated concepts [10, 11]. On the other hand, from the perspective of human-centered design (HCD), an in-depth empathic understanding of people should be developed before designers can generate solutions rooted in people's actual needs [12]. Recently, data-driven approaches towards user study have been utilizing user-generated content (e.g., online customer review, social media data) to extract and analyze users' opinion in a very large scale [13, 14]. However, these methods tend to focus on gathering explicit knowledge about people's current and past experiences, rather than learning their potential future experiences through delving deeper into their desires and dreams [15-17]. Thus, researchers have been emphasizing the critical role of empathy in human-centered design for the in-depth and comprehensive understanding of users' needs [18-20].

According to The Field Guide to Human-Centered Design by IDEO [12], empathy in design is defined as "the capacity to step into other people's shoes, to understand their lives, and start to solve problems from their perspectives". By immersing themselves in the others' world and getting closer to their lives and experiences, designers are more likely to reveal their emotions and desires and discover underlying tacit knowledge and latent needs [15, 18]. Discovering and addressing such knowledge and needs is crucial for improving user experience and facilitating market success. Recognizing the vital role that empathy plays in the design process, many researchers have been exploring methods to support the empathic design process (e.g., design probes [21], co-creation [22, 23]). However, to date, developing empathy is largely dependent on the designer's empathic capacity, which is subjective and intuitive [18, 24, 25]. Thus, discovering design opportunities through empathic design is often serendipitous. Yet, to the best of our knowledge, an effective computer-aided approach to facilitate empathic design has not been found in the literature.

In recent years, we have seen the concept of artificial or computational empathy being introduced in different disciplines, including AI [26, 27], social robotics [28, 29] (with applications in education [30] and healthcare [31]), and marketing [32]. Artificial empathy, as a new area of interdisciplinary research, focuses on modeling human empathy computationally into artificial agents. To the best of our knowledge, there has been no attempt to introduce or discuss artificial empathy in the context of human-centered design. Therefore, this paper aims to develop a framework of artificial empathy for human-centered design based on the synthesis of cutting-edge insights from several relevant fields of study. By decomposing the subjective and complex concept of empathy, we identify the modules and components of artificial empathy that can be modeled computationally. We discuss the potential roles of artificial empathy in human-centered design, introduce our proposed framework, and identify future research opportunities.

## 2  A Multidisciplinary Synthesis

Our framework for artificial empathy for HCD (to be introduced in section 3) draws on the synthesis of three literature strands, as depicted in Figure 1. In the following, we review the three literature streams.

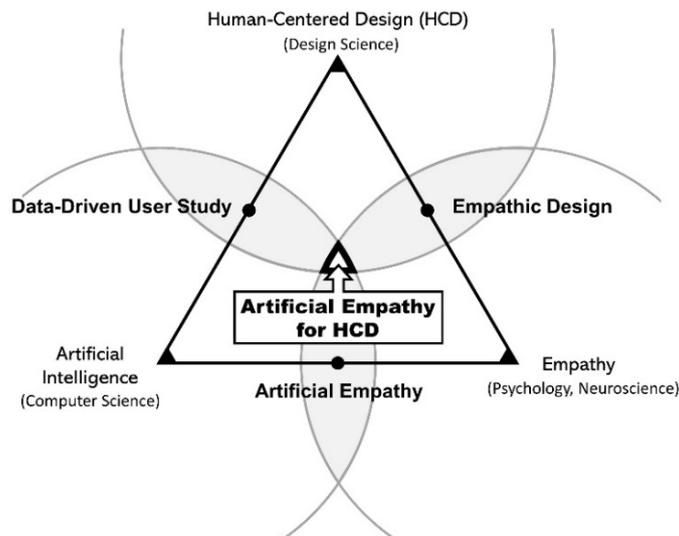

Figure 1: Artificial empathy for HCD draws on human-centered design, empathy, and AI research

## 2.1 Data-Driven User Studies

The employment of AI techniques for user studies has been drawing increasing interest in design as well as marketing research [13, 14, 33]. The expansion of online user-generated data has escalated the big-data analytics of such data using AI. This field of research has advanced the mining of valuable insights into users' thoughts, preferences, and opinions on a very large scale. Depending on users' awareness of creating the data, user-generated data can be classified into two categories: user-generated content and user-generated behavior [34].

User-generated content, which has been a widely utilized online data source for user studies, refers to the public information that users intentionally create. Commonly used user-generated content includes online reviews on e-commerce websites and social media data. Due to the massive amount, ease of access, and relatively structured data of user-generated content, they have been used in a variety of user study tasks, usually with natural language processing (NLP). The NLP tasks that leverage user-generated content include needs extraction [35], sentiment analysis [36], user profiling [37], usage context extraction [38], and emotion recognition [39]. In addition, researchers in the AI community have also introduced multimodal approaches for analyzing user-generated content beyond just textual data. For example, Poria et al. [40] developed a model based on long short-term memory network (LSTM) to capture contextual information and perform multimodal sentiment analysis using user-generated videos. Notably, recent research in engineering design [41] leveraged both text and image in user-generated content to extract product usage context and sentiments toward product features.

However, user-generated content has inherent biases. People with extreme opinions (either positive or negative) are more likely to leave their comments than those with moderate opinions, resulting in an underreporting bias [42]. The analytic results can thus be biased toward the extremes. Likewise, social media data has been shown to have a population bias, e.g., most Reddit users in the US are male and young in age [43]. Thus, the user understanding developed based on such data may not be representative of the larger population.

On the other hand, user-generated behavior (also known as the digital footprint [4]) is the information that is generated as a result of user actions. Widely used approaches include clickstream analysis and web analytics (e.g., Google Analytics). They have been used for tasks such as user behavior analysis [44], identifying trending product characteristics [45], and deciding innovation strategy [4]. However, besides the commonly concerned privacy issue [34], another limitation of user-generated behavior is that it can only be generated through using interactive systems. Thus, for broader fields of design (e.g., engineering design), user behavior data may be difficult to obtain in similar ways.

## 2.2 Empathic Design

In human-centered design, empathic design entails developing a deep and comprehensive understanding of people's circumstances and experiences in order to foster empathy and uncover insights [18, 20]. Design researchers tend to consider empathy as a type of knowledge, and thus empathic understanding as a form of knowledge construction [20, 46]. According to the literature [15, 16], designers may learn from people in three different ways: by listening to what they say, observing what they do and use, and discovering what they know, feel, and dream. The different ways lead to different levels of knowledge about people.

By listening to what people express and think, we learn the explicit knowledge of user opinions [15]. Conventional approaches to gathering user opinions include questionnaires, interviews, and focus groups. The above-introduced applications of NLP techniques to analyze user-generated content also lie at this level of understanding. By observing what people do and use, researchers learn the actual behaviors of users in their natural habitat, which leads to the observable knowledge of people and their context [15, 16]. In human-centered design, the term context refers to "all factors that influence the experience of a product use" [16]. Conventional approaches to observable knowledge are contextual inquiry [47] and ethnographic observation [48]. They are both field data-gathering methods that involve in-depth observation, interview, and documentation to investigate the users' work practices and behavior.

Finally, by discovering what people know, feel, and dream, researchers can reach a deeper understanding and learn tacit knowledge, which could reveal even deeper understanding of latent needs, i.e., needs not recognizable until the future [15, 16]. However, such depth of understanding can be difficult to develop because the internal feelings and mental states of people can hardly be sensed directly. Popularly, people's feelings can also be accessed through self-assessment surveys, which ask participants to report their emotions based on predetermined emotion categories [49] or to report their psychological feelings based on semantic differential scales [50].

However, getting to know people's affective feelings is not enough. A great portion of empathic understanding comes from the cognitive aspect, which depends on the perspective-taking and inference ability of designers [18, 46]. Such cognitive understanding can be inferred from explicit and observable knowledge by paying careful attention to various clues to unfold underlying patterns [46]. However, the ability of perspective-taking needs training to develop and can vary from person to person [24]. Existing design methods supporting perspective-taking include role-playing [51], where the designers act out others' lives and experiences, as well as simulating experiences physically (e.g., simulating visual impairment by limiting vision [19]) or virtually (e.g., social perspective-taking through virtual reality [52]).

Furthermore, the empathic understanding in design research is usually recognized as an interplay of both affective and cognitive aspects [20, 46, 53, 54]. This affective-cognitive interplay involves feeling others' emotions and trying to take their perspective and make sense of them [18, 20]. Designers often make use of their own knowledge about the world and their experiences, blending with the observed information of others to achieve empathic understanding [55, 56]. This emphasizes that developing empathic understanding is not just about studying the stakeholders, but it is also dependent on more general knowledge and experience (i.e., experiential knowledge [56]) of the designers. However, it is also argued that designers should not bring their own prejudices and stereotypes about others when trying to develop empathy [46]. Thus, it can be crucial to determine the right level of background knowledge that designers should bring with them.

In addition, researchers have emphasized the role of imagination in empathic design [18, 25, 53, 57]. In this paper, by "imagining," we do not refer to the perspective-taking process typically described as "imagining oneself in others' situation" [18]. Instead, we refer to the generative mental process that comes up with potentially better alternatives (imagined situations) [23, 25] and assesses the perceived usefulness (imagined use) of these alternatives [53, 57]. Designers who cannot vigilantly imagine alternatives can get lost reflecting on others' experiences without producing any new insights [25].
In typical co-design sessions, participants are provided with generative toolkits to create artifacts by themselves and tell a story about what they made [16, 22, 58]. The purpose of these activities is to facilitate or trigger people's imagination and expressions about what they want to experience in the future [25]. On the other hand, designers' imagination is generated in their minds based on an already developed understanding of people. Designers should vigilantly seek potential design opportunities that lead to a more imaginative future.

As Norman [59] emphasized: "*One cannot evaluate an innovation by asking potential customers for their views. This requires people to imagine something they have no experience with.*" Designers must find a way to represent their imagination in ways that people can understand, feel, and imagine. Besides sketching and prototyping (e.g., mockups and paper prototyping), designers can also leverage storytelling techniques to convey their imagination to people and between designers [60]. By creating a future with new experiences and inviting people to have a glance at it, their imagination can be triggered, and their thoughts and feelings about the new situation can be learned. Thus, storytelling can become a platform for collaborative imagination [25].

## 2.3 Artificial Empathy

The idea of enabling empathic interactions between AI and ordinary people has led to the concept of artificial empathy, which aims to provide AI with human-like empathic capacity [29]. So far, the development of artificial empathy in AI and robotics has been based on a strong foundation of psychology and neuroscience. From a theoretical perspective, empathy can be modeled by two distinct categories: affective empathy and cognitive empathy [61, 62]. This view is also widely accepted in human-centered design [18, 53].

Affective empathy is the automatic response and mimicking of others' affective states, which is based on the mirror neuron system from a neuroscience perspective [63, 64]. For example, when observing others in pain, the mirror neurons are activated as if the observers are experiencing pain themselves, thus developing shared feelings. This automatic responding process is also known as emotional contagion [63]. Other researchers suggest that mirror neurons are only responsible for forming an internal representation of the observed states and require insula to associate the internal representation with the observer's state [65, 66]. On the other hand, cognitive empathy is the ability to understand others' mental state (beliefs, desires) by taking their perspective [67], which is related to theory of mind [68, 69].

Another view of modeling empathy takes an evolutionary and developmental perspective. For instance, the Russian Doll model of empathy [70] argues that empathy is developed from low-level unconscious emotion contagion to high-level cognitive mechanisms. Following the Russian Doll model, Asada [28, 29] proposed a developmental model of self-other recognition as a part of Cognitive Developmental Robotics. The developmental perspective of empathy modeling is widely adapted in artificial empathy [26-29, 32].

Following the developmental model of artificial empathy proposed by Yalcin and DiPaola [26, 27], the components forming the low-level mechanisms of artificial empathy include emotion recognition, emotion representation, and emotion expression [26, 27]. Emotion recognition has been extensively investigated in affective computing [71, 72]. This area of research focuses on recognizing people's emotions from various modalities of information, including visual (e.g., facial expression, body gesture), audio (e.g., audio recording), and textual (e.g., customer review, social media). Apart from the unimodal approaches that use a single modality for emotion recognition, researchers have also been fusing more than one modality for multimodal emotion recognition [73], introducing physiological modalities [74], and considering contextual information encoded in images or videos [40]. For adding the emotion expression capacity, Asada [29] showed examples that recognize and express emotions in body gestures and facial expressions, while in both examples, the internal emotion states are represented by vector embeddings. More recently, Casas et al. [75] developed an empathic conversational AI that generates empathic responses by fine-tuning a GPT model.

Additionally, researchers have proposed an emotion regulation module related to the self-other distinction capability of the AI agent. With this module, the AI is supposed to modify and regulate the "raw" emotion representation based on the agent's mood and personality, as well as its relational and social link to the person being empathized with [26, 27, 76].

On the other hand, the components that form the high-level cognitive mechanisms of artificial empathy include the appraisal theory and theory of mind [26, 27]. Appraisal theory is the cognitive mechanism of emotion that states that our emotions are triggered by our appraisal of the events and situations in our environment [62]. This mechanism gives rise to affective perspective-taking, i.e., taking others' perspective and understanding their feelings and the causes of their feelings [68]. Theory of mind refers to one's ability to attribute mental states (e.g., desires, intentions, and beliefs) to others [69]. This mechanism is also referred to as cognitive perspective-taking, or as the ability to infer the beliefs of others [77]. However, the computational modeling of the cognitive mechanisms can be challenging. So far, Rabinowitz et al. [78] used meta-learning to model theory of mind computationally, and the model managed to recognize other agents' false beliefs. Jara-Ettinger [79], on the other hand, employed inverse reinforcement learning (IRL), which aims to learn a reward function (goal inference) through observing the behaviors of an agent. A recent study [80] showed that the ability of theory of mind emerged in the latest versions of GPTs, even if this ability was not explicitly engineered into these large language models.

The mechanism behind the emergence of theory of mind in GPT is yet unknown. The model could have discovered language patterns behind theory of mind that are unknown to humans or learned this ability spontaneously from the training data [80]. In either way, the massive amount of general human knowledge it was trained on seems to be the key. This can be associated with the theory-theorists' view of cognitive empathy, which suggests we infer others' minds based on our internal storage of abstract and generalized knowledge about the world and about ourselves and others [81].

## 3 Artificial Empathy for Human-Centered Design (HCD)

### 3.1 The Role of Artificial Empathy in HCD

Based on the foregoing literature review, we conclude that there are three major differences between the potential development of artificial empathy for HCD and that in other disciplines:

(1) *More human-centered, less human-like.* The main body of artificial empathy research has been focused on the human-like empathic ability of AI and empathic interaction between humans and AI agents. However, the role of empathy in human-centered design, as mentioned at the beginning of this paper, is to "step into other people's shoes, to understand their lives, and start to solve problems from their perspectives" [12]. This heavily depends on the construction of empathic understanding of others. Thus, artificial empathy in this scenario will focus on the interaction between people and their context, rather than people with the AI agent.

(2) *Ability to imagine and engage collaborative imagination.* The empathizer in HCD needs to continuously imagine alternatives that could lead to a potentially better situation and engage people in the imagination process [25]. This level of empathic ability is not seen in any other research area that the authors have investigated. We argue that this ability of imagining is at a higher level than the cognitive mechanisms of empathy and should happen after perspective-taking takes place.

(3) *Expressing insights, rather than emotions.* In robotics and affective computing, researchers have focused on the AI's ability to express appropriate emotions to people, either by parallelly imitating others' emotions or reactively appraising their situation [82]. On the other hand, artificial empathy in HCD should inform valuable insights or engage and elicit human expression and imagination. This requires the agent to express the contents in ways that people can easily understand, feel, and imagine.

However, despite the differences, empathy in HCD can largely overlap with artificial empathy in terms of their internal mechanisms and representations. Therefore, the development of artificial empathy for HCD can draw references and inspirations from the more developed research areas.

Overall, we argue that the potential role of artificial empathy in HCD should be as assistants and tools for designers that facilitate the empathic understanding process, rather than as empathic companions.

### 3.2 The Framework of Artificial Empathy for HCD

With the role of artificial empathy in mind, and by combining insights from different research areas, we propose the framework of artificial empathy for human-centered design. The aim of this framework is to break down the relatively subjective and complex concept of empathy in HCD into different modules and components that can be modeled computationally. Drawing an analogy from human empathic intelligence, we illustrate the framework in Figure 2, which has three essential modules: 'senses', 'mind', and 'expression'.

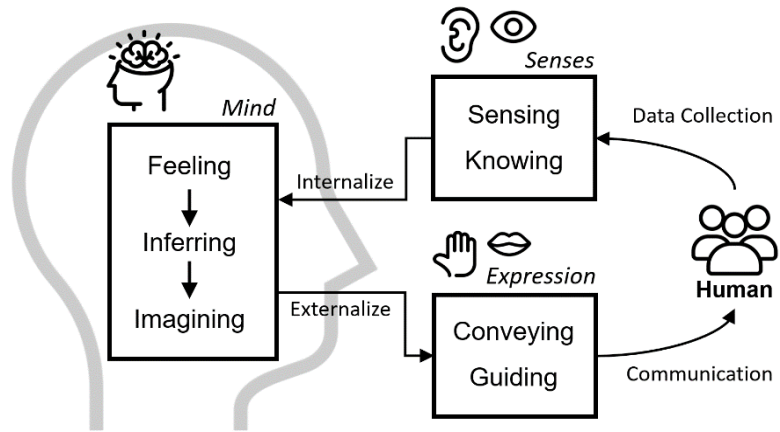

Figure 2: The framework of artificial empathy for HCD

### 3.2.1 'Senses' of Artificial Empathy

To develop empathy for people, the first step is getting to know them. 'Senses' stands for the data gathering (sensing component) and processing (knowing component) ability of an artificial empathy agent to sense and get to know people and their context externally. The sensible data can come in three types:

*Verbal data*. Textual data collected from surveys, online reviews, or social media, as well as audio data collected from interview recordings or extracted from user-generated videos, can be used to extract information about people's thoughts and opinions. So far, gathering user-generated content and using NLP to analyze the text information is a rather attractive approach in design research, because it requires little expense and travel, and the statistical results based on the large scale of data look convincing and promising. In addition, it has been shown that the textual data of user feedback could also contain contextual information, and thus has been utilized to construct a knowledge graph of product usage context [38].

*Visual data*. Visual data, such as images or videos recorded during contextual inquiry, can provide information about people's physical characteristics, behaviors, and contexts. The multimodal analysis of user-generated content could potentially offer us many insights into the relationship between people and their context. Specifically, the information embedded in images and videos that users have created and uploaded online is valuable for extracting behavioral and contextual information, as well as people's physical characteristics, such as facial expressions and body gestures. So far, the engineering design community has explored extracting context information from images through object detection [41]. In the computer vision field, recent studies have delved into video analysis for object tracking [83] and human action detection in complex real-world events [84]. HCD research could benefit from adopting such techniques in analyzing videos generated by users or recorded during contextual inquiries. By doing so, we can expect to obtain more comprehensive explicit and observable knowledge about people.

*Physiological data*. There is another line of research focusing on measuring people's physiological signals using sensor technologies introduced from neuroscience. The collected physiological data can then be mapped into emotional states. Such approaches have widely been introduced in neuromarketing [85, 86]. Popular sensing techniques include facial encoding, skin conductance, electroencephalography (EEG) brain scans, and functional magnetic resonance imaging (fMRI). The gathered data reveals the deeper psychological reactions of people that can hardly be detected otherwise. While this approach has also been used to investigate the neurocognition of designers during design activities [87], we still lack a clear understanding of what these measurements may indicate for empathic design and how to use them properly [20, 88]. The potential use of physiological data for artificial empathy needs to be clarified in future research efforts.

By sensing and processing data directly, the artificial empathy system will acquire comprehensive knowledge about people at the explicit and observable levels, which also contains clues for tacit knowledge. This module of "senses" corresponds to the unconscious sensing ability of humans, without the need for purposefully trying to make sense of the information. Developing empathic understanding, from the perspective of knowledge construction, is a hierarchy from gathering explicit and observable knowledge to the inference and imagination of tacit knowledge and latent needs. One cannot effectively take the perspective of others and develop empathy if they barely know them explicitly.

### 3.2.2 'Mind' of Artificial Empathy

The information gathered by 'senses' is then internalized by the 'mind' of artificial empathy. This module corresponds to the empathic mental processes conceptualized by Surma-Aho and Hölttä-Otto [20], which is an internal aspect of empathic understanding.

In this framework, we follow the developmental view of empathy modelling that the higher-level mechanisms of empathy are developed upon lower-level mechanisms. In fact, researchers suggest that the internal representation of low-level empathy can be used to inform high-level empathy [26]. Thus, we identified three levels of components in the 'mind' of artificial empathy: feeling, inferring, and imagining.

At the *feeling* component, we refer to the unconscious process of emotion recognition and internalization. For humans, this ability largely depends on the mechanism of mirror neuron system [63, 64]. Emotion recognition, as we introduced earlier, has been intensively investigated in affective computing and Neuromarketing. Researchers have been utilizing a variety of data modalities, including verbal, visual, and physiological data to develop more accurate recognition systems. In fact, it has been shown that multimodal approaches for computational emotion recognition can have an edge over human capability [89]. In HCD, researchers have explored emotion recognition through facial encoding to measure if shared feelings have been established between designers and users [90]. However, the recognized emotions of people are not always consistent with their real feelings [90], especially during non-contextual interviews where people may jokily talk about their pain points.

At the *inferring* component, we consider the cognitive mechanisms of empathy modeled by Yalcin and DiPaola [27], which include the appraisal theory to understand people's affective states (emotion, feeling), and theory of mind to understand their mental states (intention, belief, and desire). Based on the knowledge known to the artificial empathy system, the attempt of inferring the causes and intensity of an emotion as well as inferring the belief and desire of people would be performed. Recent research efforts in computer science have investigated NLP for emotion inference from the perspective of appraisal theory [91, 92]. The researchers created a corpus of event descriptions and annotated emotion labels by both the experiencer and human inferrers, and tested NLP's performance of inferring emotions based on given events. The emotion inference through appraisal theory differs from the feeling component by the cognitive process involved. For example, typical emotion recognition tasks involve texts with emotional expressions and tones that we can feel directly. On the other hand, appraisal theory aims to build connection between an event and its experiencer's emotion. The event description text can be as neutral as "*I bought my own horse with my own money I had worked hard to afford*" (example from Troiano et al. [92]), but we can take the perspective of the experiencer in the event and infer his/her joyful feeling. However, in HCD practices the contexts of events can be more complex and involve interactions with detailed product components, and the situated application of NLP based appraisal theory is yet to be investigated.

On the other hand, theory of mind considers the inference of mental states from explicit and observable knowledge. In psychology and computer science, the ability of theory of mind of human or AI agents is often tested through false belief tasks [78, 80, 93], where the test subjects are asked to predict the behavior of another person or agent who has a false belief about a situation. However, design researchers argued that the false belief tasks have validity limitations for HCD context and proposed to use empathic accuracy test instead to measure designer's thought inference accuracy [94]. As mentioned in section 2.3, the concept of theory of mind has been studied in AI research using meta-learning [78] and inverse reinforcement learning [79]. But these attempts took virtual agents in simulation environments and thus the current stage of this line of research is not applicable to HCD. Large language models (e.g., GPT), on the other hand, can make inference of people's mind based on textual descriptions of real-world events and situations [80] and are more promising to work in an HCD scenario. But again, the performance of such AI-based mind inference will need to be validated specifically for HCD.

In the proposed framework, the component of *inferring* depends heavily on the explicit and observable knowledge formed in the previous process, as well as a comprehensive background knowledge about the common senses of the world and the common reasonings of people's minds. The combination of these knowledge sources resembles the user knowledge construction process in design, i.e., blending epistemic knowledge about people with the experiential knowledge of the designers [56]. The inferred understanding should then better lead to the very nature of the problems that people are experiencing [25].

In HCD practice, having an in-depth understanding of the underlying causes of people's positive or negative emotions as well as their unspoken intentions, beliefs, and desires, designers are then ready to imagine alternatives to improve the current situations. At the *imagining* component, we refer to the AI's ability to "twist the reality [25]", i.e., the ability to imagine alternatives for current situation toward a potentially better future situation, without which the system could stuck in reflecting on the available knowledge without producing meaningful insights. For example, before noise-cancelling earphones came into market, the need for such technology could hardly be found from explicit knowledge because people took it as a commonsense property that earphones can hardly be used in noisy places. However, by removing a potential cause (external noise) of negative emotions (annoying) appraised from the situation (public places), artificial empathy can potentially imagine alternatives like "earphones that eliminate external noise". Apart from emotions, imagining could also consider people's unmet mental states like intentions and motivations. For example, when using a mirror, people's motivation is likely to be "to see myself exactly as others see me", but the indoor lighting can hinder the fulfillment of that motivation. By introducing a new entity (natural lighting) to the current situation (e.g., indoor dressing table) that is necessary for the fulfillment of motivation, artificial empathy could imagine alternatives like "mirror that provides natural lighting". These are two examples that we expect the imagining component to work, and it depends heavily on the empathic understanding developed through pervious components.

With *imagining* capacity, artificial empathy will acquire the capability to reach the deepest level of empathic understanding, i.e., latent needs, which are the needs that are not recognizable by people until the future [15]. Thus, by definition, they can hardly be gathered or extracted externally from available data. Prior work by Zhou et al. [95] has studied NLP based latent needs finding through exploring extraordinary use cases, which can be considered as a form of reality twists that imagine alternative use cases for a design target (e.g., alternative usage contexts or alternative target users). However, the needs elicited by this approach are limited to use-case-related ones and are based on explicit knowledge without the empathic understanding of people. On the other hand, the component of imagining in the proposed framework will require the internal representation of the empathic understanding knowledge constructed by previous components combined with the commonsense knowledge and reasoning shared across modules and components of the system.

### 3.2.3 'Expression' of Artificial Empathy

The "expression" module requires the ability of artificial empathy to externalize the empathic understanding formed in the "mind" module in ways that effectively facilitate communication with stakeholders. Thus, the "expression" module will require a chatbot-like communication system with the ability to externalize knowledge and provide guidance.

The first component of "expression" is *conveying*, which represents the externalization of empathic understanding for both designers and stakeholders to understand. This component aims to represent knowledge and insights in ways that people can easily understand, feel, and imagine. For explicit, observable, and tacit knowledge of empathic understanding, *conveying* can be performed by directly organizing the internal representation of the knowledge into natural language. However, as most stakeholders are not trained to imagine things that do not yet exist like designers do [59], the interpretation and articulation of latent needs can be supported with storytelling [60]. In general practices, generative storytelling has been shown to be technically feasible through natural language generation [96], and it can even be combined with text-to-image generation AI to facilitate the creation of a more immersive experience [97]. With the recent advances of large language models [98] and diffusion-based models [99], we can expect the employment of generative storytelling or storyboarding in HCD soon.

The other component of "expression" considers communication and iteration. In typical participatory design practices, designers need to continuously communicate with stakeholders throughout the process, providing guidance and receiving feedback to improve empathic understanding. Likewise, we can expect that in many cases, the initially collected data will be inadequate to support the *inferring* and *imagining* processes in artificial empathy. Therefore, the *guiding* component involves intentionally asking questions and providing stimuli or inspirations to engage and encourage people's expression and imagination. More radically, it could also ask people to create simple drawings or 3D modeling through the system interface. The feedback will be gathered by "senses" and thereby starting another loop for iteration.

The "expression" module in the proposed framework does not directly contribute to empathy development. Instead, it serves as an interface between the system and the people it is learning about to confirm and validate the developed empathic understanding and engage these people into collaborative imagination to gain more meaningful insights. The two modules, including "senses" and "expression," together form the external aspect of empathic understanding [20].

### 3.3 Artificial Empathy as Multimodal and Multidimensional Knowledge Construction

As mentioned earlier, we identify empathy in HCD as a form of knowledge construction [20, 46]. Artificial empathy in this context aims to develop in-depth and comprehensive knowledge about the empathic understanding of stakeholders involved in the product or service to be designed.

According to our framework, such knowledge is constructed level-by-level. First, through 'senses', the system obtains explicit knowledge of opinions and thoughts, as well as the observable knowledge of behaviors and contextual information. Second, 'mind' recognizes and understands the affective states and infers the motivation, beliefs, desires of people, which together form tacit knowledge. Finally, through vigilant reflection of the current situation in 'mind' and iterative communication to people with 'expression', knowledge of their latent needs can be discovered and recognized. Thus, this level-by-level knowledge construction allows not only the finding of people's actual needs, but also the reasoning behind the findings.

It is important to note that multimodality is needed for each module of artificial empathy. 'Senses' collect data from the multimodal world, 'mind' reasons based on the multimodal information, and 'expression' represents knowledge in multimodal manners.

Furthermore, the knowledge construction of artificial empathy involves multidimensional integration. Apart from the information collected by 'senses', artificial empathy also requires the integration of commonsense reasoning [100]. For example, 'senses' need taxonomic reasoning [100] to know the categories and relationships of instances. 'Mind' should comprehend the theory of action and change [100] as well as commonsense psychology [101] to infer what is happening next in the situation and in people's minds. 'Expression' also needs commonsense reasoning and knowledge about the world to perform storytelling. However, although it may be easy for humans, commonsense reasoning can be very challenging for AI to achieve [100]. Another dimension of knowledge integration is the general knowledge of project-specific background. For example, when designing medical equipment, the system should have knowledge of the functions, structures, and working principles of the equipment as well as its workflow involving the interaction with doctors or patients.

Taking inspiration from Sarica et al. [102], Figure 3 illustrates the architecture of knowledge integration in artificial empathy. To date, integrating multimodal and multidimensional knowledge in an artificial system has been challenging. However, we have seen recent large foundation models increase their capability in comprehending vast knowledge and complex reasoning. Specifically, the latest version of GPT (i.e., GPT-4) has demonstrated human-level commonsense reasoning and outperformed previous large language models [103, 104], while also having multimodal capability [103]. This shows great promise for future research on artificial empathy for HCD.

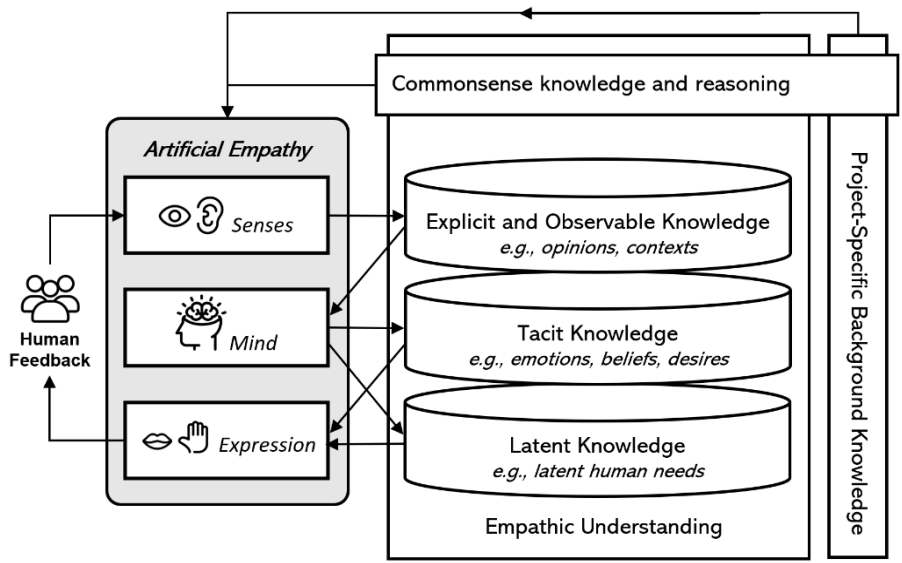

Figure 3: The architecture of knowledge integration in artificial empathy

# 4 Discussion

## 4.1 The Potential Benefits of Employing Artificial Empathy In HCD

As mentioned in the first section, empathic design is usually intuitive, and the discovery of innovation opportunities through empathic design is thus serendipitous. Through artificial empathy, we can potentially get to high-level empathic understanding built upon multidimensional knowledge and clues that human designers can hardly comprehend.

Moreover, user study is usually considered as a tradeoff between depth and scale, as Norman [17] emphasized: "*deep insights on real needs from a tiny set of people, versus broad, reliable purchasing data from a wide range and large number of people.*" By employing artificial empathy, depth and scale can possibly be handled simultaneously. By gathering large-scale and multimodal data, in-depth understanding can be developed through computationally modeled developmental empathic mechanisms. Alternatively, a web-based interactive platform of artificial empathy can be developed to involve a large number of stakeholders virtually, allowing the AI agents to interact with them and gather data concurrently.

Modern design projects often involve complex systems with a large number and variety of stakeholders that conventional participatory design practices may not accommodate. The development and employment of artificial empathy can be particularly valuable in this context.

## 4.2 Evaluating Artificial Empathy in HCD

Surma-aho and Hölttä-Otto [20] identified six categories of empathy measures in design, including empathic tendency, beliefs about empathy, emotion recognition, understanding mental content, shared feelings, and prosocial responding. However, most of these empathy measurements focus on evaluating the empathic connection between designers and people, which is not applicable for evaluating an artificial agent's empathic ability. For example, existing empathic design studies have employed Interpersonal Reactivity Index (IRI) to measure trait empathy [105, 106] and recognition of facial expression mimicry to measure affective empathy developed by designers [90]. However, an AI agent can easily answer an IRI survey with high empathic tendency or response to people with shared feelings by imitating their facial expression, but these cannot translate to a real empathic understanding.

Among these measures, performance-based emotion recognition and understanding of mental content are directly linked to empathic understanding in design processes [20, 24]. Emotion recognition measures low-level empathic understanding and has been intensively investigated in affective computing. Understanding of mental content can be assessed using empathic accuracy test [107, 108], which measures the accuracy of inferring thoughts and feelings of others. Future research is needed to validate the AI's performance of mind inference for HCD, e.g., measuring the empathic accuracy of the machine inferred states of people comparing to their self-reported states and human designers' inferred states.

Moreover, artificial empathy's capacity of needs finding will need to be tested. In a recent study, Li and Hölttä-Otto [109] used metrics of need quality and need latency to evaluate designers' performance of needs finding. Zhou et al. [110] categorized customer needs based on Kano's model [111] to identify attractive, one-dimensional, must-have, and indifferent needs. However, suitable evaluation approaches for artificial empathy's need-finding capacity will need to be developed in future research. We suggest that the AI agent's performance of both mind inference and need-finding should be compared with those of human designers under the same evaluation metrics.

## 4.3 Current Research Gaps and Opportunities

Following the proposed framework, artificial empathy can potentially provide valuable insights and human-centered innovation opportunities. However, artificial empathy can be a very complex system and requires the development and effective collaboration of different modules situated to HCD. We raise some of the current research gaps as follow and call for future research efforts:

*Multimodal machine learning for HCD*. Although multimodal machine learning has been intensively investigated in AI as well as in other aspects of engineering design [112 ,113], it has not been widely situated for HCD practices. The specific requirements of HCD should be considered when developing such systems. Based on the previous analysis of the "mind" module and the available modelling techniques for appraisal theory and theory of mind, we suggest the most promising data types of artificial empathy's internal representation can be natural languages and knowledge graph. These representations have been intensively studied in engineering design [114, 115] but haven't been seen in empathy research except in [116]. Thus, the multimodal fusion of data collected by "sense" and the mapping to these internal representations of "mind" can be a good direction to push forward the research of artificial empathy for HCD.

*Modelling cognitive mechanisms of empathy for HCD*. As mentioned in section 2.3 and section 3.2.2, the computational modellings of appraisal theory and theory of mind have been investigated in AI and robotics. However, the situated employment of such techniques in HCD to infer a deeper understanding of people has yet to be seen, e.g., how to identify design-related events and infer design-related mental states and emotions from them. Particularly, the appraisal theory has been applied in the research of emotional design [117-119]. The insights from this line of research could potentially be valuable for the development of artificial empathy. For example, Desmet [117] suggests that apart from the product (event) itself, people's concerns about the product are also essential for the cognitive appraisal of emotion. Additionally, another gap lies in the lack of baselines for measuring the performance of these cognitive mechanisms. Future works can contribute to the topic by constructing datasets of mind inferences in different design-specific contexts.

*Machine imagination and storytelling.* Recent research has shown creativity capacity of large language models performing creative design concept generation tasks [8, 9]. Machine imagination and storytelling could also benefit from generative models. However, in empathic design, instead of generating functional concepts, we need machine imagination to focus on the exploration of potential new needs, and storytelling to convey to the stakeholders a potentially beneficial future that they can feel and imagine. Thus, the generation of such contents and their performance will need to be addressed in future works.

*Potential ethical issues.* Ethical issues of AI like bias and fairness have been investigated intensively in recent years [120, 121]. According to our framework of artificial empathy, most modules will be driven by data gathered from people and the contents they create, we need to be extremely cautious not to pick up biases from the data and lead to unethical results. Moreover, artificial empathy is perceived to lack authenticity [65], e.g., does having the capabilities to perform empathic tasks as programmed necessarily translate to having real empathy? These issues may need to be investigated before the development of artificial empathy for HCD.

## 5 Conclusion

Based on an interdisciplinary investigation into research areas around user study, AI, and empathy, this paper presents a framework of artificial empathy for HCD. We have also discussed the role, benefits and ethical issues of artificial empathy in context of HCD and explicated the current research gaps to call for future research efforts. We hope this paper may stimulate more research to design and develop artificial empathy for human-centered design.